# Automatic Tunning of MapReduce Jobs using Uncertain Pattern Matching Analysis

This paper has been originally published as "A study on using uncertain time series matching algorithms for MapReduce applications" in Journal of Concurrency and Computation: Practice and Experience - Special Issue in Cloud Computing Scalability, John Wiley Publisher. We realized that the original title is not appropriate and cannot be found by people working in this area. Therefore, this text is for changing the title but the original paper can be found at the rest of this text (starting from the next page). For citation, please cite the original title as:

*NB Rizvandi, J Taheri, R Moraveji, AY Zomaya, "A study on using uncertain time series matching algorithms for MapReduce applications", Journal of Concurrency and Computation: Practice and Experience - Special Issue in Cloud Computing Scalability, John Wiley Publisher (2012)*

# A Study on Using Uncertain Time Series Matching Algorithms for MapReduce Applications

Nikzad Babaii Rizvandi[1,2], Javid Taheri[1], Albert Y. Zomaya[1], Reza Moraveji[1,2]

[1] Center for Distributed and High Performance Computing,
School of Information Technologies, University of Sydney, Sydney, Australia

[2] National ICT Australia (NICTA), Australian Technology Park

Sydney, Australia

nikzad@it.usyd.edu.au

*Abstract*—**In this paper, we study CPU utilization time patterns of several MapReduce applications. After extracting running patterns of several applications, the patterns along with their statistical information are saved in a reference database to be later used to tweak system parameters to efficiently execute future unknown applications. To achieve this goal, CPU utilization patterns of new applications along with its statistical information are compared with the already known ones in the reference database to find/predict their most probable execution patterns. Because of different pattern lengths, the Dynamic Time Warping (DTW) is utilized for such comparison; a statistical analysis is then applied to DTWs' outcomes to select the most suitable candidates. Furthermore, under a hypothesis, we also proposed another algorithm to classify applications under similar CPU utilization patterns. Finally, dependency between minimum distance/maximum similarity of applications and their scalability (in both input size and number of virtual nodes) are studied. Here, we used widely used applications (WordCount, Distributed Grep, and Terasort) as well as an Exim Mainlog parsing application to evaluate our hypothesis in automatic tweaking MapReduce configuration parameters in executing similar applications scalable on both size of input data and number of virtual nodes. Results are very promising and showed the effectiveness of our approach on a private cloud with up to 25 virtual nodes.**

I. INTRODUCTION

Recently, businesses have started using MapReduce as a popular computation framework for processing large-scaled data in both public and private clouds; e.g., many Internet endeavors are already deploying MapReduce platforms to analyze their core businesses by mining their produced data [1-3]. Therefore, there is a significant benefit to application developers in understanding performance trade-offs in MapReduce-style computations in order to better utilize their computational resources [4].

MapReduce users typically run a few number of applications for a long time. For example, Facebook, which is based on Hadoop (Apache implementation of MapReduce in Java), is using MapReduce to read its daily produced log files and filter database information depending on the incoming queries; such applications are repeated millions of times per day in Facebook. Another example is Yahoo where around 80-90% of their jobs are based on Hadoop [5]. The typical applications here are searching among large quantities of data, indexing the documents and returning appropriate information to incoming queries. Similar to Facebook, these applications are run million times per day for different purposes.

One of the major problems with direct influence on MapReduce performance is tweaking/tuning the effective configuration parameters [6] (e.g., number of mappers, number of reducers, etc) for efficient execution of an application when scaled to the size of input file or number of nodes. These optimal values not only are very hard to properly set, but also can significantly change from one application to another. Furthermore, obtaining these optimal values usually needs running an application for several times with different configuration parameters values: a very time consuming and costly procedure. Therefore, it is very important to find the optimal values for these parameters before actual running of such application on MapReduce platforms.

The work presented in this paper is an attempt for solving the problem of automatic tweaking/tuning of MapReduce configuration parameters (here, the number of mappers and number of reducers) for efficient execution of an application on cloud. This is achieved by predicting uncertain CPU utilization pattern of new applications based on the already known ones in a database. Then we show how it scales with respect to input size, and number of virtual nodes.  More specifically, we propose a two-phase approach to extract patterns and find statistical similarity in uncertain CPU utilization patterns of MapReduce applications. In the first phase, profiling, few applications are run on several small-sized input data files with different sets of MapReduce configuration parameters (here number of mappers, and number of reducers) to collect their execution/utilization profiles in a Linux environment. Upon obtaining such information –the CPU utilization time series of these applications–, their statistical information at each point is calculated. These uncertain CPU utilization values are then stored in a reference database to be later used in the second phase, i.e., matching. To use our approach for automatic tweaking of MapReduce configuration parameters, when a new unknown application is submitted, it is first run on the same small-sized input data and with the same sets of configuration parameters. Then in the matching phase, a pattern-matching algorithm is deployed to find minimum distance (maximum similarity) between stored CPU utilization profiles and the new application. If the new application is fairly similar to an application in the reference database –for the same values of configuration parameters and same sets of small-sized input data–, then it can be concluded that these two applications will probably have the same CPU behavioral for large-sized input data files (scalability in input data size) as well. As a result, if the optimal values of configuration parameters (here number of mappers and reducers) of the application in the reference database were already calculated, it is very likely that these values also result in optimal (or at least fairly close sub-optimal) settings for the new application on the same input data size. Among the two phases of the approach (profiling and matching), profiling of an experiment is a time consuming part. In this paper, we executed $8 \times 8 \times 5 (= 320)$ experiments for each application where the number of mappers and reducers were 4,8,12,16,20,24, 28, or, 32; the size of our input data was 5GB, 10GB, 15GB, or, 20GB; and, each experiment were run ten times to collect its statistical information. The time required for profiling changes from application to application. For example, the total profiling time for processing of 5G of WordCount data on 10 virtual nodes is 202 hours and 40 minutes while for Exim Mainlog Parsing on the same input data size and number of virtual nodes, the total time is 69 hours and 20 minutes.

To demonstrate our approach, Section II highlights the related works in this area. Section III provides some theoretical background for pattern matching in uncertain time series. Section IV explains our approach in which pattern matching is used to predict behavior of unknown applications. Section V details our experimental setup to gauge efficiency of our approach and introduces a hypothesis to classify applications. Discussion and analysis is presented in Section V, followed by conclusion in Section VI.

## II. RELATED WORKS

Early works on analyzing/improving MapReduce performance started around 2005; such as an approach introduced by Zaharia et al [7] addressed the problem of improving the performance of Hadoop for heterogeneous environments. Their approach was based on the critical assumption in Hadoop for homogeneous cluster nodes that tasks progress linearly. Hadoop utilizes these assumptions to efficiently schedule tasks and (re)execute the stragglers. Their work introduced a new scheduling policy to overcome these assumptions. Besides their work, there are many other approaches to enhance or analysis the performance of different parts of MapReduce frameworks, particularly in scheduling [8], energy efficiency [4, 9, 10] and workload optimization[11]. A statistics-driven workload modeling was introduced in [10] to effectively evaluate design decisions in scaling, configuration and scheduling. The framework in [10] was utilized to make appropriate suggestions to improve the energy efficiency of MapReduce applications. A modeling method was proposed in [9] for finding the total execution time of a MapReduce application. It used Kernel Canonical Correlation Analysis to obtain the correlation between the performance feature vectors extracted from MapReduce job logs, map time, reduce time, and, total execution time. These features were acknowledged as critical characteristics for establishing any scheduling decisions. Recent works in [12, 13] reported a basic model for MapReduce computation utilizations. Here, at first, the map and reduce phases were modeled using dynamic linear programming independently; then, these phases were combined to build a global optimal strategy for MapReduce scheduling and resource allocation. In [14-17], linear regression is applied to model the total number of CPU tick clocks/execution time of an application needs to execute and four MapReduce configuration parameters. These configuration parameters are: number of Mappers, number of Reducers, size of file system and size of input file.

The second part of our approach in this work is inspired by another discipline (Speaker recognition) in which similarity of objects is also the center of attention and therefore very important. In speaker recognition (or signature verification) applications, it has been already validated that if two voices (or signatures) are significantly similar – based on a same set of parameters as well as their combinations –; then, they are most probably produced by a unique person [18]. Inspired by this well proved fact, our proposed technique in this paper hypothesizes the same logic with the idea of pattern feature extraction and matching, an area which is widely used in pattern recognition, sequence matching in bio-informatics and machine vision. Here, we extract the CPU utilization pattern of unknown/new MapReduce applications for a small amount of data (not the whole data) and compare its results with already known patterns in a reference database to find similarity. Such similarity will show how much an application is similar to another application. As a result, the optimal values of configuration parameters (here number of mappers, and number of reducers) for unknown/new applications can be set based on the already calculated optimal values for the known similar applications in the database.

The last part of the work is to study the scalability behavioral of our algorithms for cloud environments. Scalability in cloud is generally divided into two groups: horizontal scalability and vertical scalability [19]. Horizontal scalability is the ability of an application to be scaled up to meet demands through replication; e.g., through distribution of requests across a pool of servers. Such scalability addresses the traditional load balanced models with respect to integrality of components in a cloud-computing environment. Vertical scalability, on the other hand, is the ability of an application to scale under load so that it can maintain its performance while the number of concurrent requests is increased. Although generic load balancing solutions can certainly assist in optimizing environments where application need to scale up by reducing their overheads, such solutions cannot solve core problems that prevent vertical scalability. In cloud platforms in particular, overhead can negatively impact performance (such as TCP session management, SSL operations, and compression/caching functionality) for the whole system.

In [20], the problem of service applications scalabilities on different clouds have been studied. For example, in Infrastructure-as-a-Service (IaaS) clouds –like Amazon EC2 that mainly works based on virtual machine (VM) technology– VMs can scale on both horizontally (by adding more service replicas) and vertically (by redefining and increasing a single VM resources). IaaS clouds can also be scaled vertically by adding more clusters or network resources. However, IaaS scalability is still too service-level oriented to be automatically managed; i.e., scaling decisions are made on the basis of pure infrastructure metrics and thus users' involvements is crucial for its success. As a solution, full automation and application of scalability rules (or load profile-based models) to holistically control services are granted for future developments on top of IaaS Clouds. Although these advanced high-level management automation capabilities lay close to the aforementioned PaaS features, they only deal with deployment and runtime service lifecycle stages. As a result, scaling applications in cloud environments still faces some "old-fashioned" challenges. It means, detecting code parallelism –could be offered as a PaaS–, and distributing application components in clusters and service operations in multi-core architectures will receive massive research interest to efficiently fulfil scalability requirement of future cloud applications. Another comprehensive research is performed in 2011 to study scalability issues in clouds [21]. This study (1) pointed out several critical issues about automatic scaling of applications on cloud environments, (2) presented the related state-of-the-art efforts, and, (3) highlighted existing challenges.

The horizontal scaling of MapReduce applications on public cloud like Amazon Elastic MapReduce is restricted to the Cloud resources and is provided at an additional cost [22]. Therefore, it becomes important to execute an application with large-size input data effectively on the provided Cloud resources. As a MapReduce application behavioral has strong dependency on values of the configuration parameters [23, 24], automatic effective tweaking of these parameters by Cloud load balancer results in effective usage of the Cloud resources and therefore save money and time. Our approach, in this work, is an attempt for enabling a Cloud load balancer to automatically tune a given MapReduce application parameters by comparing it to previously executed MapReduce applications in a reference database. We will also study the scalability of our approach when affected by the similarity of MapReduce applications, the size of input data files, and, the number of virtual nodes.

## III. THEORITICAL BACKGROUND

Pattern matching is a well-known approach—particularly in pattern recognition—to transform a time series pattern into a mathematical space. Such transformation is essential to extract the most suitable running features of an application before comparing it with reference applications in a database to find its similar pairs. Such approaches have two general phases: (1) profiling phase, and (2) matching phase. In the profiling phase, the time series patterns of several applications are extracted. Several mathematical operations are then applied on these patterns (including magnitude normalization); results are stored in a database to be used as references for the matching phase. In matching phase, the same procedure is repeated for an unknown/new application first; and then, the time series of this application are compared with those stored in the database –using a pattern matching algorithm– to find the most similar ones.

### A. Uncertain time series

A time series $\varphi_c(.)$ is called **c**ertain time series when its data values are fixed/certain: $\varphi_c(.) = [\varphi_c[1], ..., \varphi_c[N]]$ where $\varphi_c[i]$ in the value of time series at time $i$. A time series $\varphi_u(.) = [\varphi_u[1], ..., \varphi_u[N]]$ is called **u**ncertain when there is uncertainty in its data values [25] and be formulated as:

$$\varphi_u[i] = \varphi_c[i] + e_\varphi[i]$$

where $e_\varphi[i]$ is the amount of error/uncertainty in $i^{th}$ data point. Due to uncertainty, the value of each point is considered as independent random variable with statistical mean ($\mu_\varphi[i]$) and standard deviation ($\sigma_\varphi[i]$). These values are calculated during analyzing time series in the profiling phase.

In MapReduce application, because the length of CPU utilization time series as well as their values in each point may change for even several identical executional environments of an application –i.e., the same input data file size, number of mappers, and number of reducers–, we considered the CPU utilization time series "uncertain" and try to use its statistical information in our similarity measurements.

### B. Pattern matching

Similarity measurement algorithms have been frequently used in pattern matching, classification and sequence alignment in bio-informatics. The measurement of similarity between two (normalized) uncertain time series means to find a function: $SIM(\varphi_u, \phi_u)$ where $\varphi_u(.)$ and $\phi_u(.)$ are two time series without the same length. This function is typically designed as $0 \leq SIM(\varphi_u, \phi_u) \leq 1$, where greater values means higher similarities. In this case, $SIM(\varphi_u, \phi_u) = 1$ should be obtained for identical series only, and, $SIM(\varphi_u, \phi_u) = 0$ should reflect no similarity at all. To this end, "similarity distance is defined as a specific distance between two uncertain time series to reflect the level of their similarity.

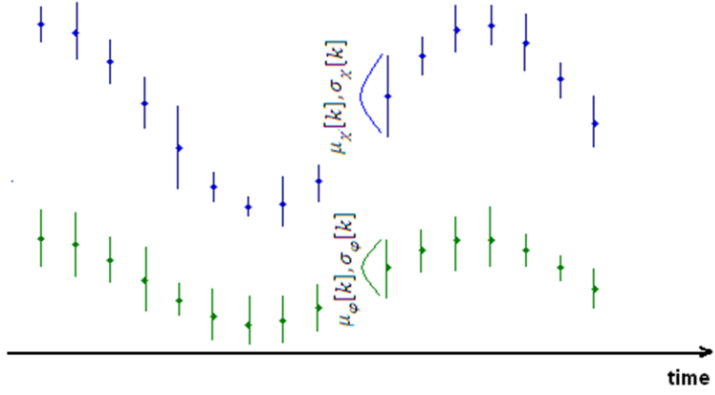

Figure 1. The distance between two uncertain time series and normal distribution of uncertainty in the k-th points.

1. Dynamic Time Warping (DTW)

DTW is among the common techniques to calculate distance/similarity between two certain time series of different lengths. This approach cannot be used to find similarity between two uncertain time series it usually results in unacceptable outcomes. DTW uses a nonlinear search to map corresponding samples of each series to find such similarity. The following recursive operation shows how such distance/similarity between two certain time series $\varphi_c(.) = [\varphi_c[1], ..., \varphi_c[N]]$ and $\chi_c(.) = [\chi_c[1], ..., \chi_c[M]]$ ($N \geq M$) is computed:

$$D(i,j) = \begin{cases} D(i, j-1) \\ D(i-1, j) \\ D(i-1, j-1) \end{cases} + d(\varphi_c[i], \chi_c[j]) \quad (1)$$

where $d(.,.)$ is the Euclidean distance between corresponding points in each series and,

$$d(\varphi_c[i], \chi_c[j]) = \|CPU(\varphi_c[i]) - CPU(\chi_c[j])\|$$

where $CPU(\varphi_c[i])$ is the value of CPU utilization at time $i$ in $\varphi_c$.

Results of these formulation is the $D[\varphi_c, \chi_c]$ matrix in which each element, $D(i,j)$, reflects the minimum distance between $[\varphi_c[1], \chi_c[1]]$ to $[\varphi_c[i], \chi_c[j]]$. As a result, $D(N, M)$ would reflects the distance/similarity between $\varphi_c$ and $\chi_c$. In this case, $\varphi'_c$ and $\chi'_c$ with equal length can always be made from $\varphi_c$ and $\chi_c$, respectively, so that $\chi'_c[i]$ is aligned with $\varphi'_c[i]$. $\varphi'_c$ and $\chi'_c$ are always made from $\varphi_c$ and $\chi_c$, respectively, by repeating some of their elements –based on $D[\varphi_c, \chi_c]$.

As mentioned earlier, because DTW cannot be directly used for uncertain time series, we only use it to produce temporary time series $(\varphi'_c, \chi'_c)$ with the same lengths. This procedure is then proceeded by applying DTW on the certain time series ($\varphi_c = mean(\varphi_u)$, $\chi_c = mean(\chi_u)$) parts of the two uncertain time series $(\varphi_u, \chi_u)$ as follows:

$$\varphi_u[i] = \underbrace{mean(\varphi_u[i])}_{\varphi_c[i]} + \underbrace{var(\varphi_u[i])}_{e_\varphi[i]}, 1 \leq i \leq N \text{ and}$$

$$\chi_u[j] = \underbrace{mean(\chi_u[j])}_{\chi_c[j]} + \underbrace{var(\chi_u[j])}_{e_\chi[j]}, 1 \leq j \leq M,$$

where $N \geq M$.

Then, to calculate the Euclidian distance between two uncertain but same length time series:

$$\varphi'_u[i] = \underbrace{mean(\varphi'_u[i])}_{\varphi'_c[i]} + \underbrace{var(\varphi'_u[i])}_{e_{\varphi'}[j]}, 1 \leq j \leq R \text{ and}$$

$$\chi'_u[j] = \underbrace{mean(\chi'_u[j])}_{\chi'_c[j]} + \underbrace{var(\chi'_u[j])}_{e_{\chi'}[j]}, 1 \leq j \leq R$$

where $R \geq N, M$.

As a result:

$$[\varphi'_c, \chi'_c] = DTW(\varphi_c, \chi_c) \quad (2)$$

It is worth nothing that DTW in this paper is only utilized to provide same length data series for $\varphi_u$ and $\chi_u$ –and not to provide their actual similarity–, because DTW does not affect statistical mean and variance of points in the two uncertain time series. Thus, if DTW maps $\varphi'_c[i]$ to $\varphi_c[j]$, then

$$\begin{cases} mean(\varphi'_c[i]) = mean(\varphi_c[j]) \\ var(\varphi'_c[i]) = var(\varphi_c[j]) \end{cases}$$

*2. Similarity measurement*

Despite many solutions that the distance between two time series is calculated by summing up distances from points aligned by DTW procedure, here, we use the square Euclidian distance to calculate such similarity. To be more specific, in our approach, two certain time series $\varphi'_c$ and $\chi'_c$ are similar when the square Euclidian distance between them is less than a distance threshold ($r$):

$$DST(\varphi'_c, \chi'_c) = \sum_{i=1}^{N} (\varphi'_c[i] - \chi'_c[i])^2 \leq r$$

For uncertain time series $\varphi'_u$ and $\chi'_u$ –where the problem is not straightforward as before–, similarity is defined as [25]:

$$\Pr\left(DST(\varphi'_u, \chi'_u) = \sum_{i=1}^{N} D^2[i] \leq r\right) \geq \tau \quad (3)$$

where $D[i]$ is a random variable equal to $\varphi_u[i] - \chi'_u[i]$. Therefore, two uncertain time series are assumed similar when probability of their Euclidian distance is more than a pre-defined threshold ($0 < \tau \leq 1$).

Because $\varphi'_u[i]$ and $\chi'_u[i]$ are independent random variables (figure 1), both $D[i]$ and $DST(\varphi'_c, \chi'_c)$ are also independent random variables. Therefore, if $<\mu_{\varphi'}[i], \sigma_{\varphi'}[i]>$ and $<\mu_{\chi'}[i], \sigma_{\chi'}[i]>$ are <statistical mean, standard derivation> of $\varphi'_u[i]$ and $\chi'_u[i]$, respectively, according to [25], $DST(\varphi'_u, \chi'_u)$ has the following normal distribution:

$$DST(\varphi'_u, \chi'_u) \sim \mathbb{N}\left(\sum_{i=1}^{N} E(D^2[i]), \sum_{i=1}^{N} Var(D^2[i])\right) \quad (4)$$

where

$$\sum_{i=1}^{N} E(D^2[i]) = \sum_{i=1}^{N} \left[\mu_{\varphi'}^2[i] + \sigma_{\varphi'}^2[i] - 2\mu_{\varphi'}[i]\mu_{\chi'}[i]\right]$$
$$+ \sum_{i=1}^{N} \left[\mu_{\chi'}^2[i] + \sigma_{\chi'}^2[i]\right] \quad (5)$$

and

$$\sum_{i=1}^{N} Var(D^2[i]) = 4\sum_{i=1}^{N} \left[\left(\sigma_{\varphi'}^2[i] + \sigma_{\chi'}^2[i]\right)\left(\mu_{\varphi'}[i] - \mu_{\chi'}[i]\right)^2\right] \quad (6)$$

The standard normal distribution function of $DST(\varphi'_u, \chi'_u)$ can be calculated as:

$$DST_{norm}(\varphi'_u, \chi'_u) \sim \mathbb{N}(0,1) = \frac{DST(\varphi'_u, \chi'_u) - \sum_{i=1}^{N} E(D^2[i])}{\sqrt{\sum_{i=1}^{N} Var(D^2[i])}} \quad (7)$$

Thus the problem in Eqn.(3) can be rewritten as:

$$\Pr(DST_{norm}(\varphi'_u, \chi'_u) \leq r_{norm}(\varphi'_u, \chi'_u)) \geq \tau \quad (8)$$

Definition 1: $r_{boundry,norm}$ is a minimum distance bound value that finds the lower bound for the standard normal probability in Eqn.(8); i.e., [25]:

$$\Pr(DST_{norm}(\varphi'_u, \chi'_u) \leq r_{boundry,norm}) = \tau \quad (9)$$

where $r_{boundry,norm} = \sqrt{2} \times \text{erf}^{-1}(2\tau - 1)$ for standard normal distribution and $\text{erf}(.)$ is an error function obtained from statistics tables [26] when working on $DST(\varphi'_u, \chi'_u)$ instead of $DST_{norm}(\varphi'_u, \chi'_u)$:

$$r_{boundry} = \frac{r^2_{boundry,norm} - \sum_{i=1}^{N} E(D^2[i])}{\sqrt{\sum_{i=1}^{N} Var(D^2[i])}} \quad (10)$$

Definition 2:
Two uncertain time series $\varphi'_u$ and $\chi'_u$ are similar with probability more than $\tau$ [25]:

$$\Pr(DST(\varphi'_u, \chi'_u) \leq r) \geq \tau \quad (11)$$

when

$$r \geq r_{boundry} \quad (12)$$

Here, $r_{boundry}$ defines the minimum distance between two uncertain series with probability $\tau$; i.e., :

$$\Pr(DST(\varphi'_u, \chi'_u) = r_{boundry}) = \tau \quad (13)$$

Based on equations (1-13), from this point forward, we use uncertain time series only; and thus, we use $\varphi'$, $\chi'$, $\mu_{\varphi'}$, $\sigma_{\varphi'}$, $\mu_{\chi'}$ and $\sigma_{\chi'}$ instead of $\varphi'_u$, $\chi'_u$, $\mu_{\varphi'_u}$, $\sigma_{\varphi'_u}$, $\mu_{\chi'_u}$ and $\sigma_{\chi'_u}$, respectively.

*C. Problem definition*

In distributed computing systems, MapReduce has been known as a large-scale data processing or CPU intensive job [6, 27, 28]. It is also well known that CPU utilization is the most important part of running an application on MapReduce. Therefore, optimizing the amount of CPU an application needs becomes important for customers to hire enough CPU resources from cloud providers as well as for cloud providers to schedule incoming jobs properly.

Inspired by the current status of MapReduce applications and their complexity, in this paper, we try to predict CPU utilization of unknown applications through comparing them with the known one. To this end, we study the similarity between (normalized) uncertain CPU utilization time series of an incoming application with the analyzed (normalized) applications in a reference database for several small input data and different sets of configuration parameter values. If the uncertain CPU utilization time series of an unknown/new application is found to be adequately similar to uncertain CPU utilization time series of another application in database; then, it can be assumed that the CPU utilization behavior of both applications would be the same for other sets of configuration parameters values as well, especially for large input file size. This fact can be used in two ways: firstly, if the optimal values of number of mappers and number of reducers are obtained for one application (e.g., WordCount), these optimal values may lead us to optimal number of mappers and reducers of other similar applications (e.g., Distributed Grep) too; secondly, this approach allows us to properly categorize applications in several classes with the same CPU utilization behavioral patterns.

IV. PATTERN MATCHING IN MAPREDUCE APPLICATIONS

In this section, we describe our technique to find the distance/similarity between uncertain CPU utilization time series of different MapReduce applications. Our approach is consisted of two phases: profiling and matching.

```
Profiling phase
1.  For iᵗʰ application in database (φᵢ):
2.      For rᵗʰ small input data (Sizeᵣ):
3.          For jᵗʰ set of configuration parameters values (Mⱼ, Rⱼ):
4.              Counter=1
5.              Do
6.                  Run application with the jᵗʰ set of parameters on a small
                    input data(Sizeᵣ)
7.                  Capture CPU utilization Time Series with XenAPI (φᵢ,ⱼ)
8.                  Counter = Counter + 1
9.              While (Counter <= 10)
10.             For kᵗʰ point in φᵢ,ⱼ
11.                 Calculate < μ_{φᵢ}[k], σ_{φᵢ}[k] >
12.                 μ_{φᵢ} = [ μ_{φᵢ}[1],…, μ_{φᵢ}[N] ]
13.                 σ_{φᵢ} = [ σ_{φᵢ}[1],…, σ_{φᵢ}[N] ]
14.             End
15.             Save {φᵢ,(Mⱼ,Rⱼ),Sizeᵣ, μ_{φᵢ}, σ_{φᵢ}} in Reference
                database
16.         End
17.     End
18. End
```

```
Matching phase
For a new unknown application(χ):
                "Extract Statistical Information"
1.  For iᵗʰ application in database (φᵢ):
2.      For rᵗʰ small input data (Sizeᵣ):
3.          For jᵗʰ set of configuration parameters values (Mⱼ, Rⱼ):
4.              Counter=1
5.              Do
6.                  Run χ with the jᵗʰ set of parameters values parameters on
                    a small input data(Sizeᵣ)
7.                  Capture CPU utilization Time Series with XenAPI (χⱼ)
8.                  Counter = Counter + 1
9.              While (Counter <= 10)
10.             Calculate μ_{φᵢ} and μ_χ under jᵗʰ set of parameters values
11.             [ μ_{φᵢ}', μ_{χ'} ] = DTW( μ_{φᵢ} , μ_χ ) : align mean times series of
                χⱼ to mean time series of φᵢ and form new mean time
                series φᵢ' and χ'
12.             For kᵗʰ point in both new uncertain time series φᵢ' and χⱼ'
13.                 Calculate < μ_{χⱼ'}[k], σ_{χⱼ'}[k] > and < μ_{φᵢ'}[k], σ_{φᵢ'}[k] >
14.                 μ_{χⱼ'} = [ μ_{χⱼ'}[1],…, μ_{χⱼ'}[R] ] and
                    σ_{χⱼ'} = [ σ_{χⱼ'}[1],…, σ_{χⱼ'}[R] ]
15.                 μ_{φᵢ'} = [ μ_{φᵢ'}[1],…, μ_{φᵢ'}[R] ] and
                    σ_{φᵢ'} = [ σ_{φᵢ'}[1],…, σ_{φᵢ'}[R] ]
16.             End
17.             Form {χⱼ',(Mⱼ,Rⱼ),Sizeᵣ, μ_{χⱼ'}, σ_{χⱼ'}}
18.         End
19.     End
20. End
                    "Candidate Selection"
21. Set pre-defined Probability threshold (τ = 0.95)
22. For rᵗʰ small input data (Sizeᵣ):
23.     For jᵗʰ set of configuration parameters values (Mⱼ, Rⱼ):
24.         For iᵗʰ application in database (φᵢ):
25.             Calculate joint mean and variance of distance between φᵢ'
                and χⱼ' From Eqn.(5-6)
26.             Calculate r_{boundry} from Eqn.(10)
27.             < φᵢ, r_{boundry} > is added to candidature pool of χⱼ
28.         End
29.     End
30. End
In candidature pool of χⱼ, the application with lowest r_{boundry} is
chosen as the highest similar application to χⱼ
```

Figure 2. Algorithms for profiling and pattern matching phases.

## A. Profiling phase

In the profiling phase, CPU utilization time series of several MapReduce applications in database along with their statistical information is extracted. For each application, we generate a set of experiments with several small input data and two main MapReduce configuration parameters (number of mappers, and number of reducers) on a given platform.

Figure 2 shows our profiling algorithm. While running each experiment, the CPU utilization time series of the experiment in each virtual node of a cloud is gathered to build a trace to be later used as the training data –this statistic can be gathered easily in virtual node (running on Linux) with the XenAPI monitoring package. Within the system, we sample the CPU usage of the experiment in a native system from starting mappers till finishing reducers with time interval of one second. If virtual nodes are homogenous, the CPU time series of nodes for an application are assumed to be approximately similar. Thus, the final CPU time series of an application is computed by averaging CPU utilization values at each point. Because of the temporal changes, several identical experiments –i.e., same input data and configuration parameters– may result in different values in each point of the extracted CPU utilization time series. Therefore, we repeat each experiment ten times and then extract the statistical <mean, variance> of each point of the time series. It is worth noting that the completion time of these experiments were insignificantly apart from each other and thus we could safely ignore their differences. Upon completion of ten experiments, the time series with its related set of configuration parameters values as well as its normalized statistical features are stored in the reference database. This procedure was repeated for all applications we intended to profile.

## B. Matching phase

In the matching phase, the profiling is also performed for the newly submitted application and then followed by the several steps to find its distance/similarity with already known applications. As shown in Figure 2, the matching phase consists of two stages: statistical information extraction and candidate selection. In the statistical information extraction stage, CPU utilization time series of a new unknown application ($\chi$) is captured by XenAPI; then statistical <mean,variance> at each point $< \mu_\chi[k], \sigma_\chi[k] >$ of the time series are extracted under the same input data sizes and configuration parameters. Here, because the length and magnitude of the new application time series might be different from those in reference database ($\varphi_i$), we first normalize them and then use DTW to make them of equal length. Result are two new uncertain time series for each application ($\varphi_i'$ and $\chi'$) to be later analyzed for extracting their statistical information at each point $< \mu_{\varphi'}[k], \sigma_{\varphi'}[k] >$ and $< \mu_{\chi'}[k], \sigma_{\chi'}[k] >$.

In the candidate selection stage, the mathematical analysis described in Section III-B is applied to calculate the similarity between twisted version of normalized uncertain time series in database ($\varphi_i'$) as well as the new unknown application ($\chi'$).

| WordCount | Exim MainLog Parsing | | | |
|---|---|---|---|---|
| | S-1 | S-2 | S-3 | S-4 |
| S-1 | **24044** | 117017 | 94472 | 228071 |
| S-2 | 80648 | **64063** | 58351 | 138222 |
| S-3 | 79431 | *63232* | **56114** | 104255 |
| S-4 | 147014 | 83655 | 81434 | **70427** |

| Terasort | Exim MainLog Parsing | | | |
|---|---|---|---|---|
| | S-1 | S-2 | S-3 | S-4 |
| S-1 | **27400** | *65102* | 65606 | 132799 |
| S-2 | 155038 | *67293* | 68455 | *69927* |
| S-3 | 123668 | 76859 | **51876** | 76589 |
| S-4 | 166234 | 77829 | 81751 | *74693* |

| Distributed Grip | WordCount | | | |
|---|---|---|---|---|
| | S-1 | S-2 | S-3 | S-4 |
| S-1 | **21529** | 105309 | 90012 | 199451 |
| S-2 | 79965 | **62890** | 68553 | 122279 |
| S-3 | 77549 | 62949 | **54309** | 101280 |
| S-4 | 142703 | 83089 | 72987 | **70198** |

TABLE 1. A sample of the minimum distance ($r_{boundry}$) between the used applications for $\tau = 0.95$ for processing 5G of input data on 10 virtual nodes.

Consequently, based on Eqn.(13) the time series in database which gives the minimum $r_{boundry}$ for predefined Euclidian distance probability ($\tau$) are chosen as the most similar application to the new application in the candidature pool. Raising the value of probability threshold ($\tau$) will reduce the number of applications in candidature pool; and consequently, increases the similarity selection accuracy.

## V. EXPERIMENTAL RESULTS

### 1. Experimental setting

Four widely known/used applications (three text processing and one sorting) were deployed and implemented to evaluate the effectiveness of our method in this work. Figure 3 shows the structure of the private cloud we used to conduct our experiments; it has the following specifications:

- Physical H/W: includes five servers, each server was a dual-core Intel Genuine 3.00GHz with 4GB memory, 1GB cache and 250GB of shared iSCSI network drive.

- Xen cloud platform (XCP) is used for virtualization has been used on top of the physical H/W. The Xen-API [29] provides functionality in high level languages like Java, C# and Python to manage virtual machines inside XCP, measure their details performance statistics as well as live-migrate them in a private cloud environment.

Debian images are used to provide our Hadoop nodes (version 0.20.2) on our servers; each virtualized debian were set to use 1 CPU, 1GB RAM, and, 50GB of disk. The number of virtual nodes were 5, 10, 15, 20, or, 25. We used XenAPI to collect runtime CPU utilisation of these nodes on a Intel(R) Core i7 (four cores, eight logical processors, and 16GB of RAM) desktop PC to monitor/extract the CPU utilization time series of applications. Performance statistics for each experiment were collected from "running job" stage to the "job completion" stage with sampling time interval of one second. All CPU usages samples are then combined to form CPU utilization time series of an experiment. For each application we executed $8 \times 8 \times 5 (= 320)$ experiments where the number of mappers and reducers were 4,8,12,16,20,24, 28, or, 32; the size of our input data was 5GB, 10GB, 15GB, or, 20GB; and, each experiment were run ten times to collect its statistical information. Our benchmark applications were WordCount, TeraSort, Distributed Grep and Exim Mainlog parsing. These benchmarks were chosen because (1) they roughly represent a variety of MapReduce applications, and, (2) there are also used as valid MapReduce benchmarks in other approaches [14-16, 30-34].

### 2. Results and Discussion

Each of the aforementioned application is executed on several small-sized input data files with a combination of difference configuration parameters (number of mappers and reducers) to form its related CPU utilization time series.

#### a) Application similarity

Table 1 and Figure 4 indicate the minimum distance ($r_{boundry}$) between CPU utilization patterns of these four applications in our experiments (for 5G of input data on 10 virtual nodes) for Euclidian distance probability of 95% ($\tau = 0.95$). In Table 1, the lowest and the penultimate lowest minimum distance between two instances of applications are indicated with bold and bold-italic, respectively. Here $\{S-1, ..., S-8\}$ are the set of number of mappers and reducers used in the experiments. Moreover, straight lines in the figure 4 show the diagonal line and the points show the position of the minimum distance between two applications which are too close to the diagonal lines. Results in this table and figure shows that diagonal numbers of these tables are always either the lowest or the penultimate lowest of all numbers, showing that two computationally similar applications always have minimum distance when run with similar configuration parameters. Based on this observation, we designed another candidate selection algorithm in which such similarities are taken into account for selecting the best set of running parameters for one applications based on its similar peers. This new approach is detailed in Figure 5 and replaces our first attempt in Figure 2.

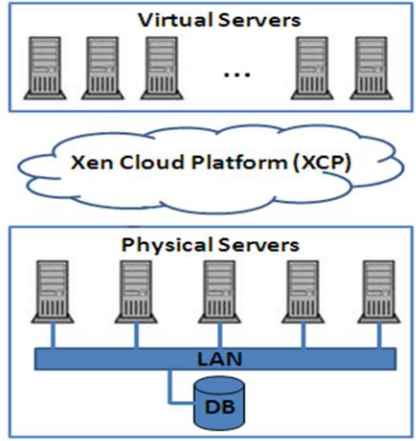

Figure 3. Overall architecture of our private cloud in conducting our experiments

Figure 5. Candidate selection and with applications pattern matching algorithm based on our application similarity hypothesis

Based on our observations, we also hypothesize that if two applications are considered "computationally similar" for short data files, they will be fairly "similar" for large data sizes too. This hypothesis can lead us to find optimal number of mappers and reducers for running a new unknown MapReduce application through first categorizing it based on its CPU utilization patterns, and then, estimating its optimal running parameters based on its peer similar applications in the dame category/class. To be more specific, assume we have N applications in our database $\varphi = \{\varphi_1, \ldots, \varphi_N\}$ along with their optimal configuration parameters.

Now, for a new unknown application $\chi$, we execute this application for the same set of input data and parameters we used to collect optimal values for $\{\varphi_1, \ldots, \varphi_N\}$; then we can chose the running parameters of $\chi$ based on the its closest application in our data base.

A direct result from table 1 indicates that WordCount, Exim and Distributed Grep can be categorized in a same CPU time series class while Terasort forms another class.

*b) Auto-similarity of applications*

To further investigate our hypothesis, we also studied the auto similarity of an application. Here, we expect that diagonal numbers for in calculating auto-similarity of all applications must be significantly larger than all other off-diagonal numbers. Table 2 shows our results and proves our point. This table in fact proves that only similar configurations parameters can produce comparatively small Euclidean distances between different experiments.

*c) $\tau$ and minimum distance $(r_{boundry})$ relation*

One of the parameters influencing minimum distance between CPU utilization time series of applications $(r_{boundry})$ is the value of Euclidian distance probability $(\tau)$. Euclidian distance probability greatly depends on the level of similarity between two applications. As expected, increasing $\tau$ always results in raising $r_{boundry}$. This observation is also well justified from a mathematical point of view as shown in Eqns. (9-10). Based on these equations, greater values of $(\tau)$ should result in greater values of $erf^{-1}(2\tau - 1)$, and consequently, greater values of the minimum distance $(r_{boundry})$ as well.

*d) Scalability for the size of the input data*

Upon finding the minimum distance between two applications for sets of parameters in Table1 and Figure 4, we investigated the relationship between the scalability of input size and the distances for $\tau = 0.95$ on 10 virtual nodes and show the results in Figure 6. Here, the input file size is 5G, 10G, 15G, or, 20G. This figure indicates that increasing size of the input data file always results in relatively greater minimum distance as well. This observation can be justified/explained by considering the fact that larger data

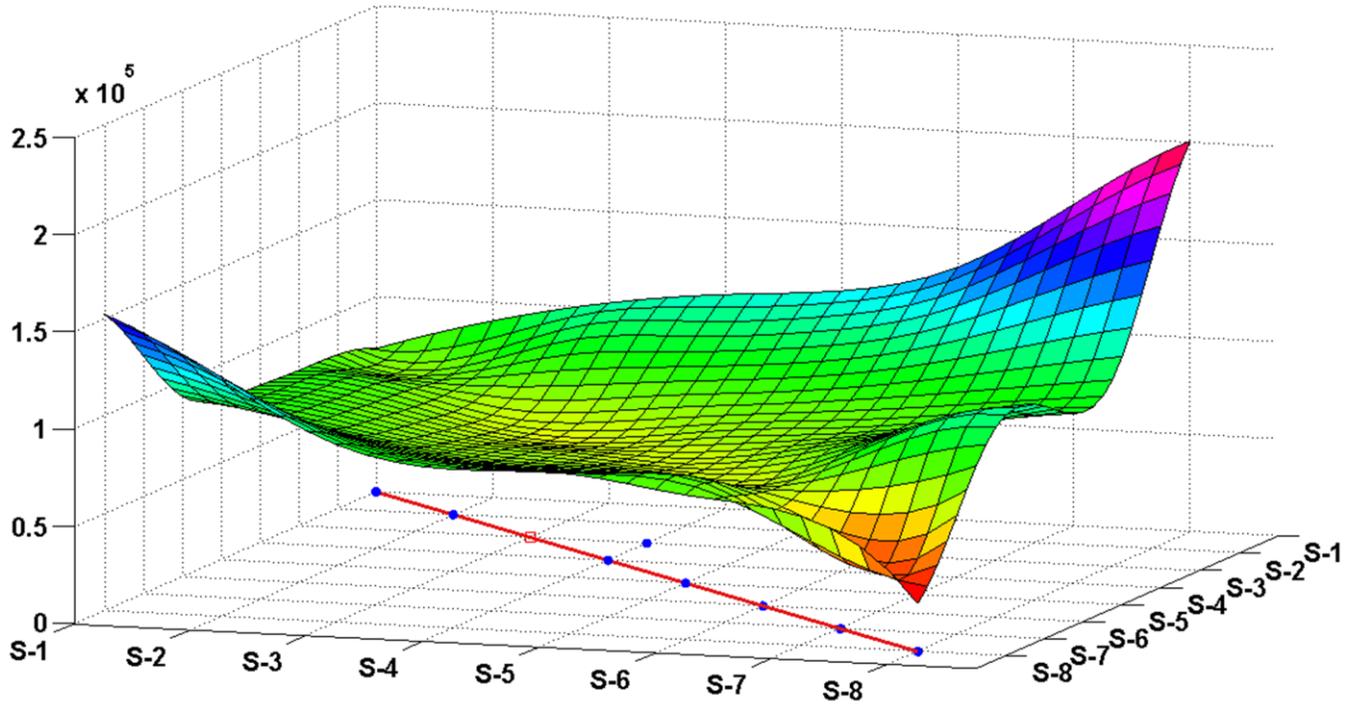

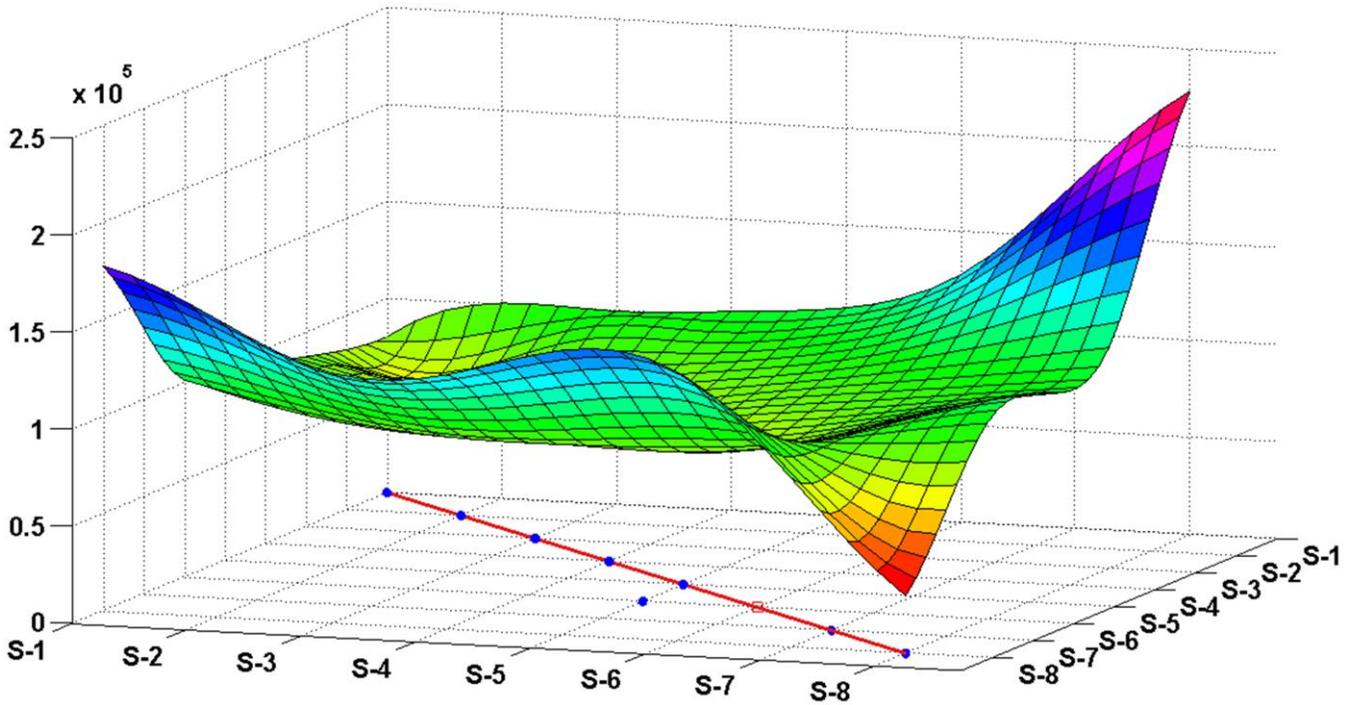

**(To be continued)**

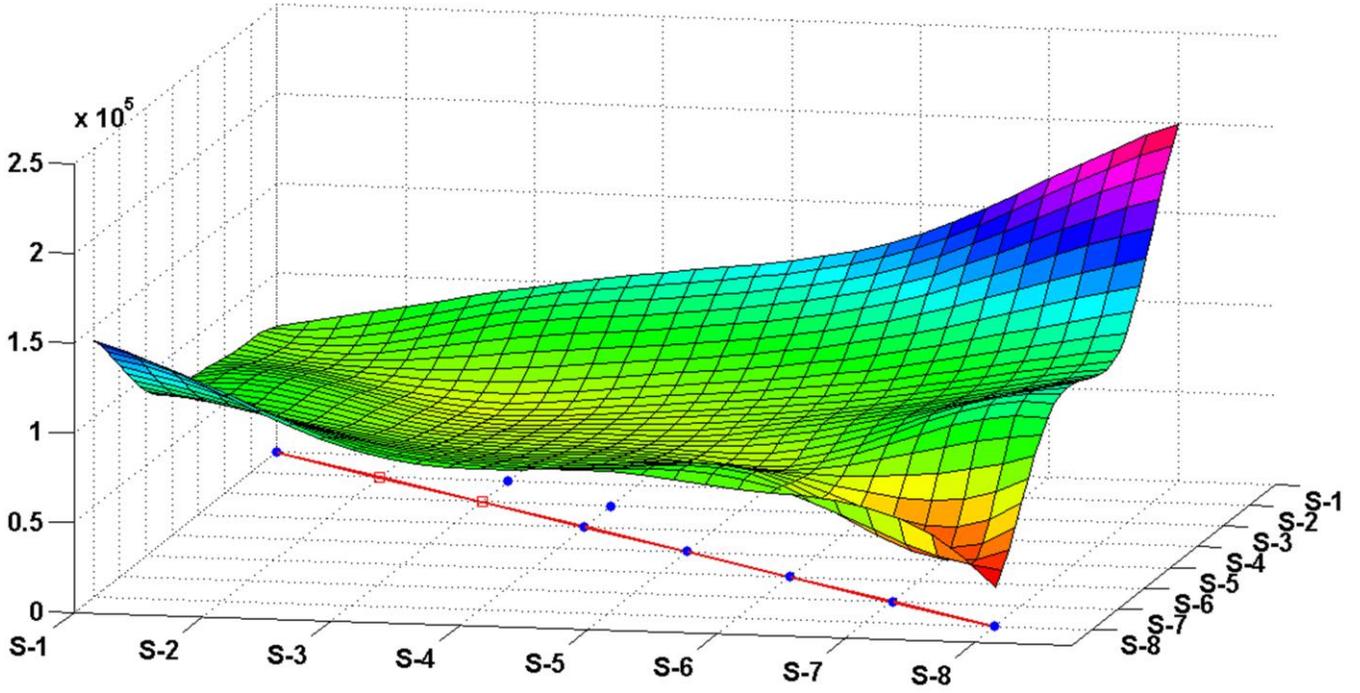
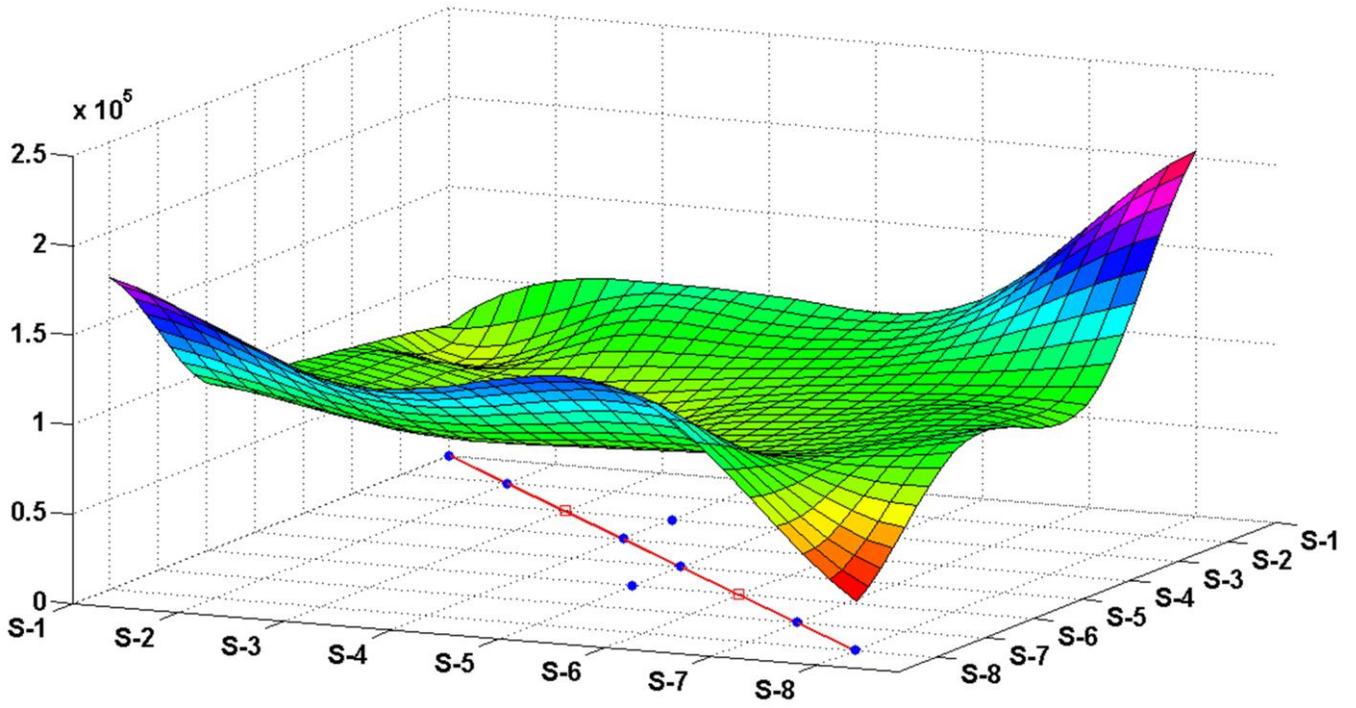

(To be continued)

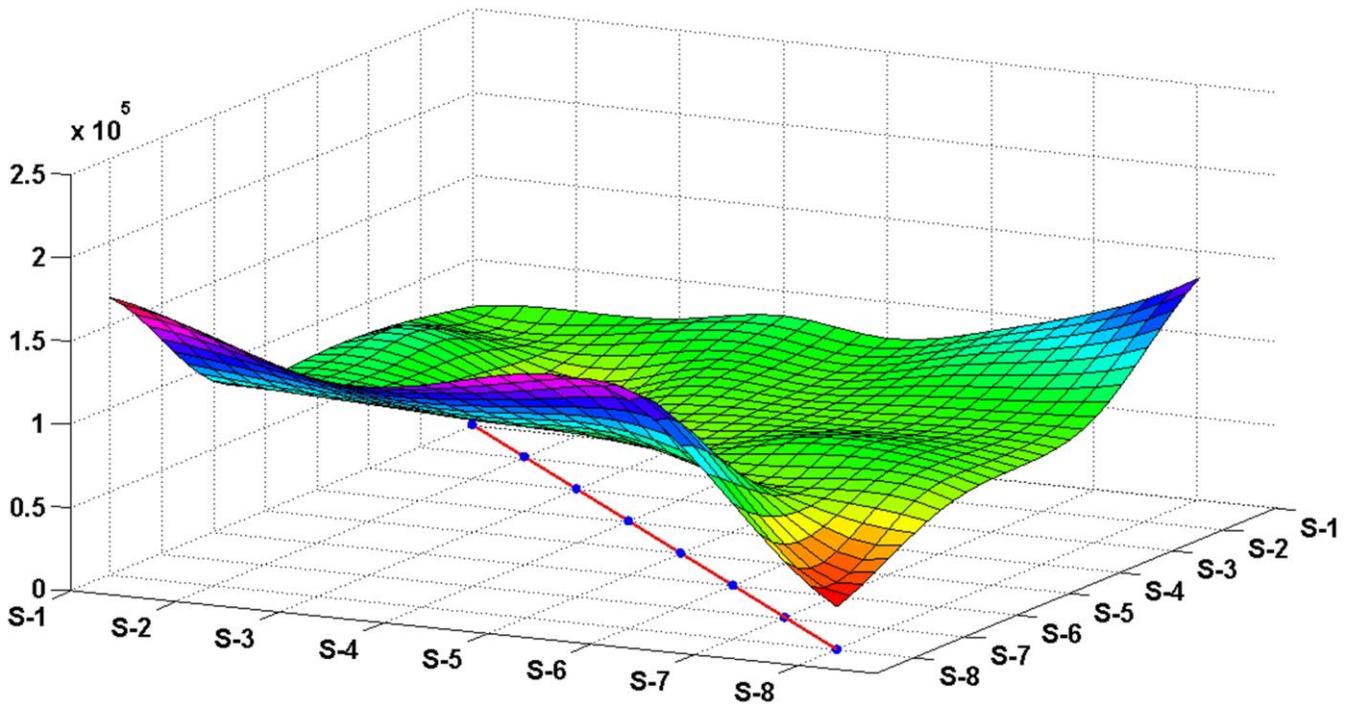
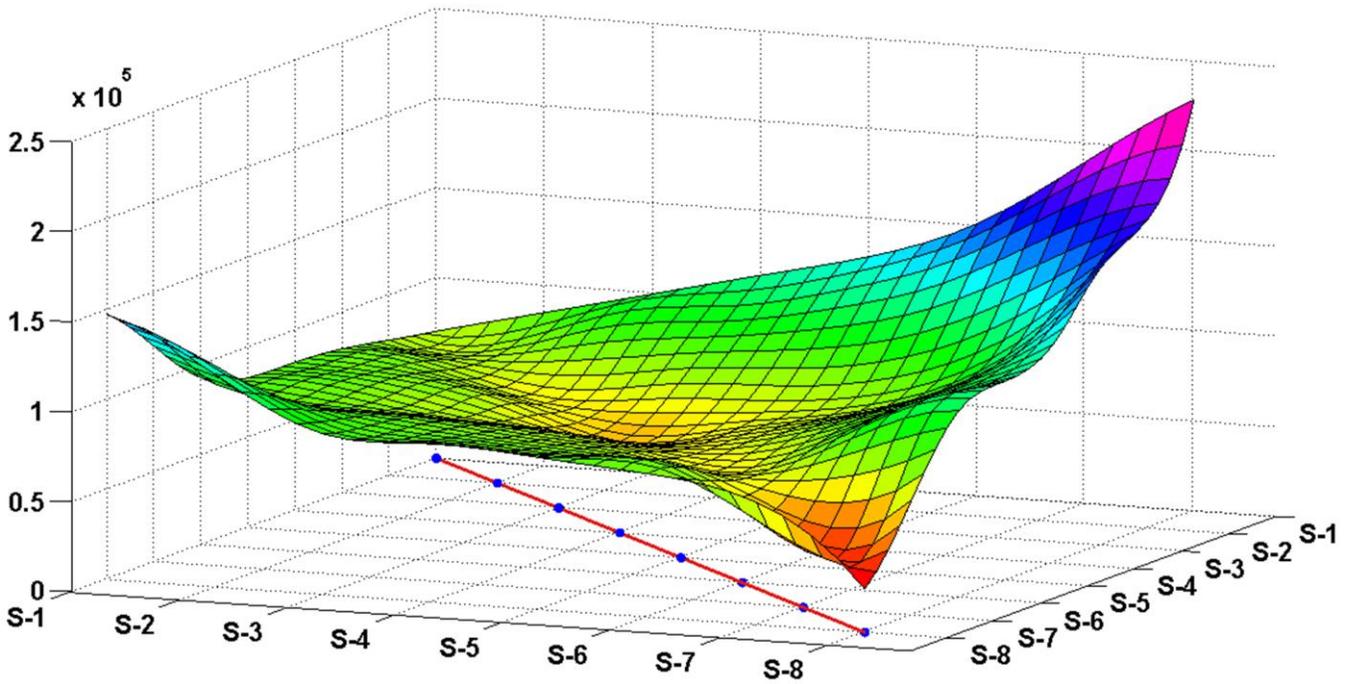

Figure 4. The minimum distance $(r_{boundry})$ between the four applications for $\tau = 0.95$ in processing 5G of input data on 10 virtual nodes. The lowest minimum distance indicates the highest similarity. The y-axis shows the value of minimum distance between applications. The straight lines show the diagonal line and the points show the position of the minimum distance between two applications. which almost close to the diagonal lines.

files always need more time to be executed as well. Therefore, the length of CPU time series will also increase and consequently causes more points to be compared.

*e) Scalability for the number of virtual nodes*

Figure 7 shows relationship between scalability of the number of nodes (5 to 25) in an experiment and the minimum distances for $\tau = 0.95$ on a 5G data file; this figure also shows that this relationship is much more complicated than the previous case. Further investigating and analysis however indicates the following facts: (1) When the number of nodes is increased –for the same data–, because the input data are mapped onto more nodes, it results shorter lengths of the CPU time series. Furthermore, because nodes are assumed identical, they are expected to produce very similar CPU time series as well. Results however show that more nodes always result in dispersed time series with more variance at each data point. This in turn causes greater levels of uncertainty, less accuracy in a system, and, greater minimum distances between experiments. (2) Increasing the number of nodes expedited execution of an application and thus results in decreasing the number of points in its CPU time series. Therefore, as mentioned in scalability for the size of input data, this also leads to lower values of minimum distances for experiments.

Comparison of Figures 6 and 7 shows that increasing either size of data or number of virtual nodes will result in greater minimum distance values. Nevertheless, as discussed, the increasing slope for scalability of virtual nodes is less than that of the size of the input data; thus, scaling the number of virtual nodes requires lower similarity threshold value compared with scaling the size of the input data. It should be also noted that although scaling the experiments in either size of data or number of nodes increases the minimum distance, our hypothesis about applications' similarities still holds. It means that scaling (in both cases) does not effect similarity among applications; mainly because, order of lowest or the penultimate lowest minimum distances among applications rarely changes when executed with different number of mappers and reducers, size of data file, and/or, number of virtual nodes.

| WordCount | | S-1 | S-2 | S-3 | S-4 |
|---|---|---|---|---|---|
| WordCount | S-1 | **1079** | 116850 | 89382 | 216474 |
| | S-2 | 116850 | **2293** | 61541 | 119652 |
| | S-3 | 89382 | 61541 | **2524** | 118166 |
| | S-4 | 216474 | 119652 | 118166 | **2980** |

(a)

| Exim | | S-1 | S-2 | S-3 | S-4 |
|---|---|---|---|---|---|
| Exim | S-1 | **325** | 63003 | 69015 | 138062 |
| | S-2 | 63003 | **987** | 42078 | 71601 |
| | S-3 | 69015 | 42078 | **1136** | 65632 |
| | S-4 | 138062 | 71601 | 65632 | **1706** |

(b)

| Terasort | | S-1 | S-2 | S-3 | S-4 |
|---|---|---|---|---|---|
| Terasort | S-1 | **361** | 135619 | 105092 | 150639 |
| | S-2 | 135619 | **1046** | 60107 | 74998 |
| | S-3 | 105092 | 60107 | **1102** | 69114 |
| | S-4 | 150639 | 74998 | 69114 | **1588** |

(c)

| Distributed Grip | | S-1 | S-2 | S-3 | S-4 |
|---|---|---|---|---|---|
| Distributed Grip | S-1 | **1895** | 106728 | 79264 | 2124544 |
| | S-2 | 111743 | **3609** | 72784 | 104756 |
| | S-3 | 79536 | 66530 | **4706** | 996546 |
| | S-4 | 207634 | 98648 | 102746 | **6098** |

(d)

Table 2. The minimum distance ($r_{boundry}$) between each application with itself for $\tau = 0.95$, input data size=5G on 10 virtual nodes.

*f) The cost of profiling and modeling*

During our experiments, we also carefully logged execution time of these algorithms. Table 3a reflects the average execution time as well as the total time for only profiling phase of a whole set of experiments for each application on 5G of input data; each application is executed 8 (possible number of mappers) times 8 (possible number of reducers) times 10 (repeating the whole experiment), i.e., 640 times in total. Table 3b shows required time for the pattern-matching phase i.e., comparing CPU time series of an application on 5G of input data to others with the same input data size in MATLAB [35], Table 3b also shows that matching TeraSort with others always takes more time; an educated guess to explain this could be related to the nature of TeraSort that is vastly different to the other applications.

*g) Future work*

Although in this work we tried to cover as many issues as possible to study the true execution behavior of known MapReduce applications and to predict the behavior of the unknown ones, we like to acknowledge that CPU utilization profile of applications alone "sometimes" does not lead to valid estimations. Therefore, the first extension to our work will be to enrich our profiling

|  | Average time for one experiment | Total time of 8*8*10 experiments per application |
| --- | --- | --- |
| **WordCount** | ~19 min | ~202 h and 40 min |
| **Exim MainLog** | ~6.5 min | ~69 h and 20 min |
| **Terasort** | ~9 min | ~96 h |
| **Distributed Grep** | ~12.2 min | ~130 h and 8 min |

(a)

|  | WordCount | Exim MainLog | Terasort | Dist. Grep |
| --- | --- | --- | --- | --- |
| **WordCount** | --- | 44 sec | 184 sec | 56 sec |
| **Exim MainLog** | 44 sec | --- | 96 sec | 64 sec |
| **Terasort** | 184 sec | 96 sec | --- | 93 sec |
| **Dist. Grep** | 56 sec | 64 sec | 93 sec | --- |

(b)

Table 3. Time required for (a) profiling and (b) pattern matching for executing the proposed algorithm for all experiments on 5G of data and on 10 virtual nodes

method by considering Disk (I/O) and Memory utilizations along with CPU utilization. This extension requires collecting three uncertain time series for each experiment in our private cloud platform. Consequently, our pattern matching procedure should also be modified to concurrently consider three uncertain times series in finding similarity among applications; this requires significantly more complex computations. It is also worth noting that considering Disk or Memory utilization can only help to better estimate behavior of applications when either a large amount data is transferred between the map and the reduce procedures, and/or, intensive computations are performed in them. Otherwise, because most disk utilization in CPU intensive MapReduce applications is performed for distributing/copying data before/after starting/finishing an application, it does not interfere with its execution time. The case for memory utilization is however more subtle as it usually provides redundant information about execution behavior of an applications; mainly because, it is tightly related to its CPU utilization. Therefore, we can confidently hypothesize that CPU utilization of applications is most probably enough to estimate execution behavior of unknown applications in most cases; nevertheless, it is always worth to investigate other possible avenues to provide more accurate solutions.

During our experiments, we also noticed that collected CPU utilization profiles of applications are sometimes too long and noisy to pass on to our DTW engine. Therefore, we also like to extend our work by extracting signaling behavior of our profiles instead of their raw data. Extracting Fourier/Wavelet coefficients applications' profiles is among our first attempts to this end. We strongly believe that using such techniques, not only will significantly reduce the size of applications' profile and expedites the whole system, but also eliminates a great amount of noise from the system to produce even more accurate estimations..

## VI. CONCLUSION

In this paper, we present a new statistical approach to find similarity among uncertain CPU utilization time series of CPU intensive MapReduce applications on a private cloud platform. Through two phases of our approach (profiling and pattern-matching), execution behavior of known applications is used to estimate behavior of the unknown ones. Profiling is performed through sampling, while our novel combination of DTW and Euclidean distance is used for pattern matching. Four applications (WordCount, Distributed grep, Exim Mainlog parsing and TeraSort) are deployed on our private cloud to experimentally evaluate performance of our approach. Results were very promising and showed how CPU utilization pattern of known application are related and therefore can be used to estimate those of the unknown ones. We also study the scalability of our approach against the size of the input data and the number of nodes in our private cloud environment.

## VII. ACKNOWLEDGMENT

The work reported in this paper is in part supported by National ICT Australia (NICTA). Professor A.Y. Zomaya's work is supported by an Australian Research Council Grant LP0884070.

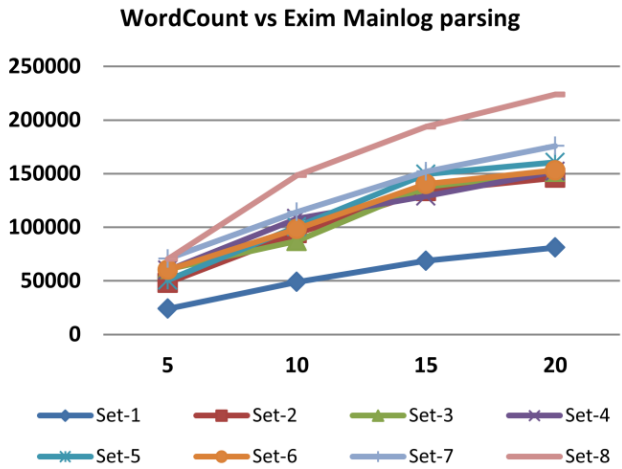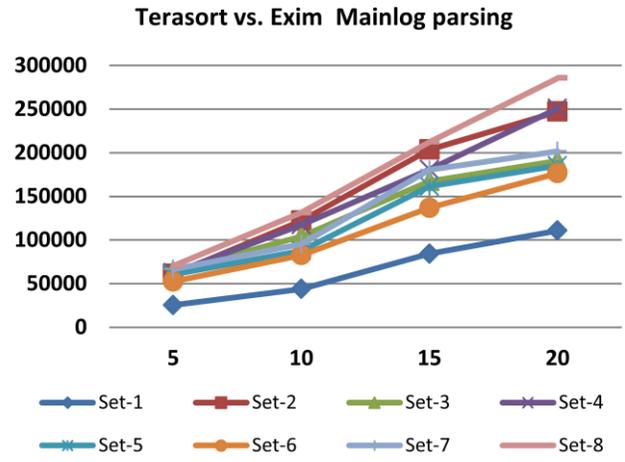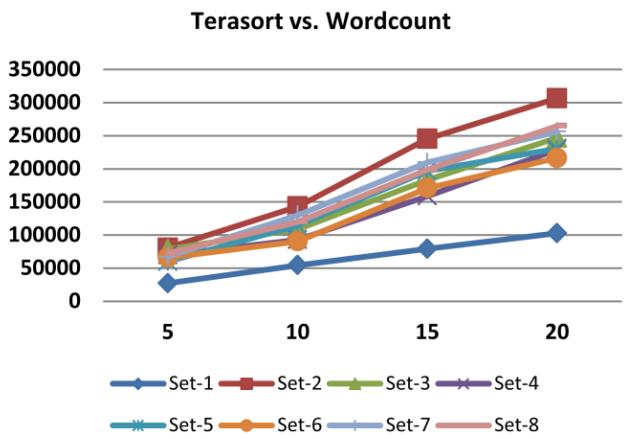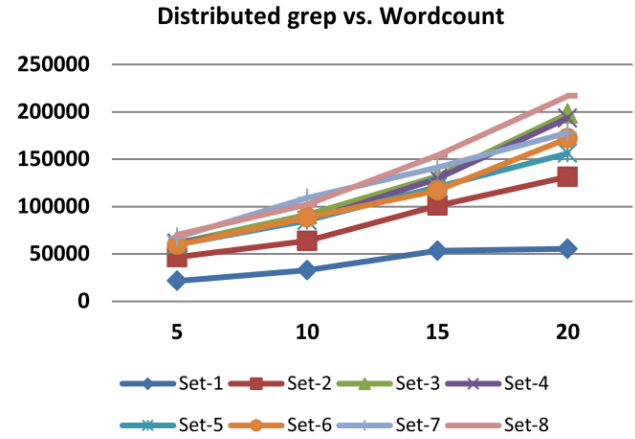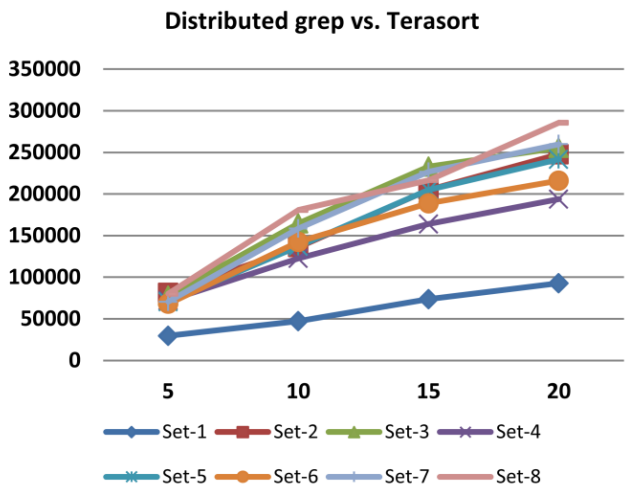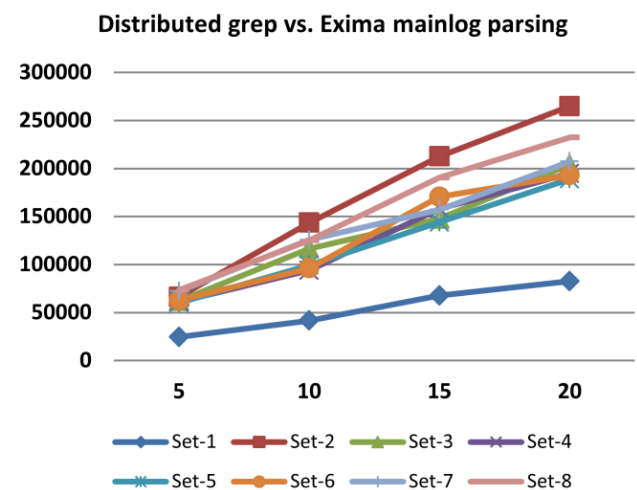

Figure 6. The dependency between the size of input file and minimum distance/maximum similarity (scalability in input file size) between applications for $\tau = 0.95,$ input data size of 5G, 10G, 15G and 20G on 10 virtual nodes. The y-axis shows the value of minimum distance between applications.

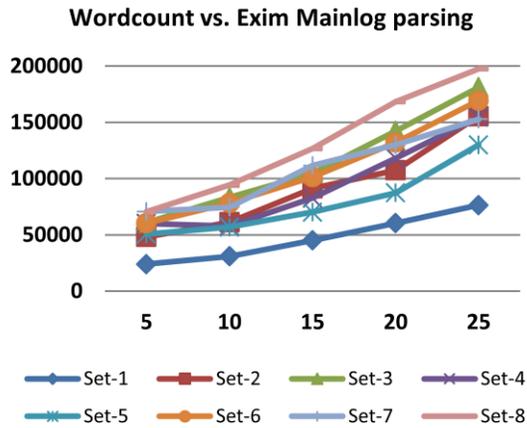
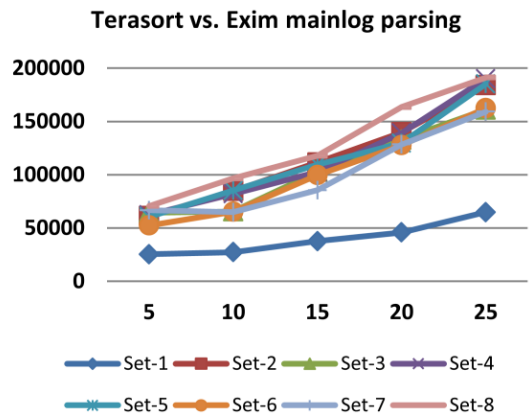
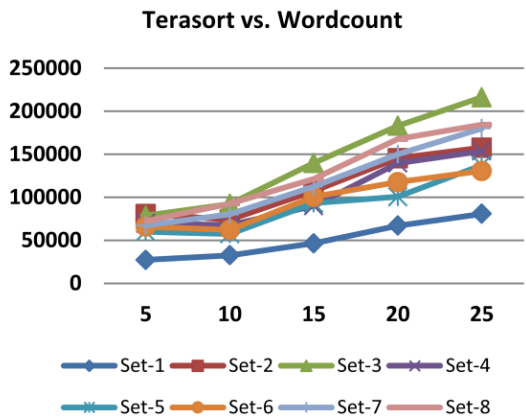
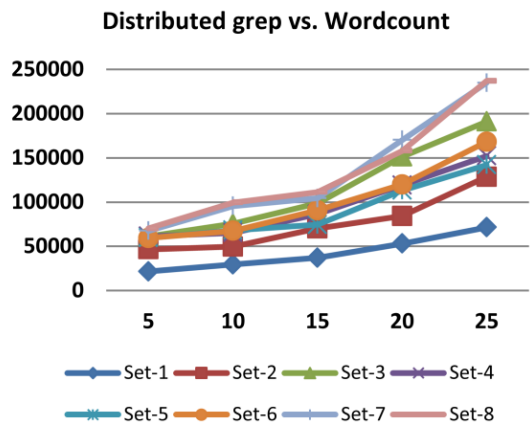
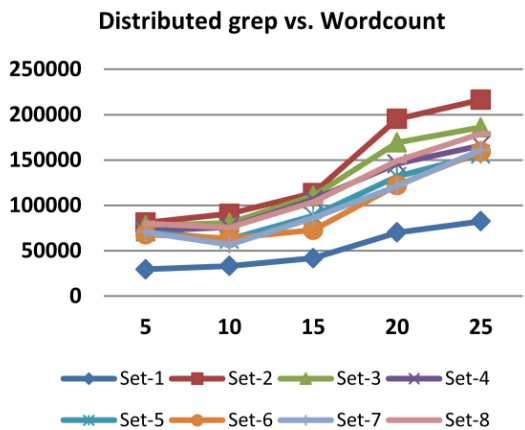
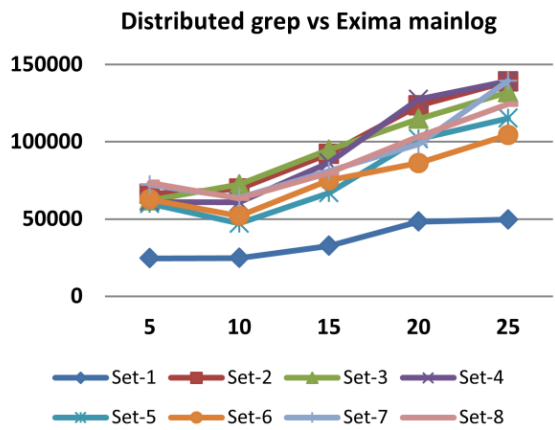

Figure 7. Dependency between the number of virtual nodes and minimum distance/maximum similarity (scalability in number of nodes) between applications for $\tau = 0.95,$ on 5, 10, 15, and 20 virtual nodes for 5G of input data size. The y-axis shows the value of minimum distance between applications.